\documentclass[conference]{IEEEtran}
\usepackage{amsmath,graphicx}
\usepackage{times}
\usepackage{epsfig}
\usepackage{graphicx}
\usepackage{amsmath}
\usepackage{amssymb}
\usepackage{algorithm}
\usepackage{algorithmic}
\usepackage{enumerate}
\usepackage{multirow}
\usepackage{multicol}
\usepackage{arydshln}
\usepackage{color}
\usepackage{amssymb}
\def\0{{\mathbf 0}}
\def\1{{\mathbf 1}}

\def\e{{\mathbf e}}

\def\p{{\mathbf p}}

\def\s{{\mathbf s}}

\def\u{{\mathbf u}}
\def\v{{\mathbf v}}

\def\x{{\mathbf x}}
\def\y{{\mathbf y}}

\def\A{{\mathbf A}}
\def\B{{\mathbf B}}
\def\C{{\mathbf C}}
\def\D{{\mathbf D}}

\def\I{{\mathbf I}}

\def\L{{\mathbf L}}

\def\P{{\mathbf P}}
\def\Q{{\mathbf Q}}

\def\S{{\mathbf S}}

\def\U{{\mathbf U}}
\def\V{{\mathbf V}}
\def\W{{\mathbf W}}
\def\X{{\mathbf X}}

\def\ie{{\textit{i.e.}}}
\def\eg{{\textit{e.g.}}}

\def\cE{{\mathcal E}}

\def\cG{{\mathcal G}}
\def\cH{{\mathcal H}}

\def\cL{{\mathcal L}}

\def\cV{{\mathcal V}}

\def\balpha{{\boldsymbol \alpha}}

\def\bLambda{{\boldsymbol \Lambda}}

\title{Complex Graph Laplacian Regularizer for Inferencing Grid States}
\author{Chinthaka Dinesh, Junfei Wang, Gene Cheung\thanks{The work of G. Cheung was supported in part by the Natural Sciences and Engineering Research
Council of Canada (NSERC) RGPIN-2019-06271, RGPAS-2019-00110.}, and Pirathayini Srikantha \\
	Emails: dineshc@yorku.ca, junfeiw@yorku.ca, genec@yorku.ca, and psrikan@yorku.ca\\
	Department of Electrical Engineering and Computer Science, York University, Toronto, ON, Canada
}
\begin{document}

\maketitle
\begin{abstract}
To maintain stable grid operations, system monitoring and control processes require the computation of grid states (e.g., voltage magnitude and angles)  at high granularity. It is necessary to infer these grid states from measurements generated by a limited number of sensors like phasor measurement units (PMUs) that can be subjected to delays and losses due to channel artefacts, and/or adversarial attacks (e.g. denial of service, jamming, etc.). 
We propose a novel graph signal processing (GSP) based algorithm to interpolate states of the entire grid from observations of a small number of grid measurements. 
It is a two-stage process, where first an underlying Hermitian graph is learnt empirically from existing grid datasets offline. 
Then, the graph is used to interpolate missing grid signal samples online in linear time. 
With our proposal, we can reconstruct grid signals with significantly smaller number of observations when compared to traditional approaches (e.g. state estimation). In contrast to existing GSP approaches, we do not require knowledge of the underlying grid structure and parameters and are able to guarantee fast spectral optimization. 
We demonstrate the computational efficacy and accuracy of our proposal via practical studies conducted on the IEEE 118 bus system.
\end{abstract}
%
%
\section{Introduction}
\label{sec:intro}
The modern electric grid is composed of intermittent generation sources and variable consumer demands that push the system to operate close to its limits. Elevated situational awareness where operators are able to rapidly infer the states of every electrical node (\ie,  bus) at highly granular intervals is necessary to ensure reliable grid operations. A costly approach would be to deploy sensors such as phasor measurement units on every bus  \cite{Mohanta2016}. 
Instead, using a limited number of sensors from which the electrical states of all nodes in the grid are inferred is a more efficient solution \cite{Nuqui2005}. Other factors that can reduce the availability of grid measurements include adversarial attacks (\eg, denial of service) and/or issues in the communication layer (\eg, delays, packet losses). Traditional state estimation processes require knowledge of the underlying grid topology and parameters. 
Also, these must be supplied with at least $2N$ sensor measurements (where $N$ is the total number of buses in the grid) that satisfy the observability criteria to allow for accurate state estimation  \cite{Monticelli2000}. 
Moreover, due to the non-linearity of the mapping between grid measurements and states, linear state estimation (\ie, DC state estimation) is widely utilized to support practical system operations. However, as numerous assumptions that include voltage magnitude of all buses being 1 p.u. are made which are not valid in practice, the estimated states will not accurately reflect the actual operation conditions of the grid \cite{Cosovic2017}.

In this paper, we leverage on the historical grid data available to infer grid states from a small subset of grid measurements in real-time using a \textit{graph signal processing} (GSP) approach \cite{ortega18ieee,cheung18}. 
Our proposal is composed of two stages. 
In the first stage, a sparse complex-valued inverse covariance matrix $\P \in \mathbb{C}^{N \times N}$ (interpreted as a graph Laplacian) is learnt from a historical grid dataset offline. 
$\P$ encodes the underlying  state inter-dependencies of the grid, without requiring knowledge of the grid structure and its parameters (\ie, bus admittance matrix). 
To learn $\P$, we extend a previous CLIME formulation~\cite{cai2011constrained} to the complex-valued case, which we transform into a linear program (LP), efficiently solvable via a state-of-the-art LP solver\footnote{A representative state-of-the-art general LP solver is \cite{jiang20}, which has complexity $\mathcal{O}(N^{2.055})$.}. 

Importantly, we construct complex-valued matrix $\P$ to be \textit{Hermitian}, which by the Spectral Theorem \cite{golub12} means that it is eigen-diagonalizable with real eigenvalues. 
This enables us to define a complex \textit{graph Laplacian regularizer} (GLR) \cite{pang17} that computes a \textit{real} value $\x^\text{H} \P \x \in \mathbb{R}$ given a complex-valued signal $\x \in \mathbb{C}^N$. 
In the second part, during online operation, we combine GLR with an $\ell_2$-norm fidelity term, forming a quadratic objective to estimate missing grid states. 
The solution is a linear system that is efficiently solved in linear time via \textit{conjugate gradient} (CG)~\cite{shewchuk94}. 


Existing works that pursue a GSP approach to interpolate missing grid states mainly differ from our proposal in how a graph variation operator is chosen to represent the grid (\eg, $\P$ matrix). 
Specifically, \cite{has2022} utilizes electrical distances to construct this matrix, \cite{Dab2023} leverages on the real components of the nodal admittance matrix, and \cite{ram21, saha2022, hasNa2022} utilize the entire complex-valued nodal admittance matrix to define the graph Laplacian matrix. 
The main issue with these techniques is that all of these require the underlying grid topology and/or parameters. 
In approaches that construct a real-valued Laplacian matrix, these either utilize only part of the admittance values or partial information about the grid (\eg, electrical distances). This limits the information embedded in the graph Laplacian about practical electrical inter-dependencies amongst the grid states. 
In other cases such as \cite{ram21}, the entire nodal admittance matrix is utilized, and although this is a complex symmetric matrix, \textit{it is not Hermitian}. 
This means that the Spectral Theorem \cite{golub12} does not apply, and the matrix is not guaranteed to be eigen-diagonalizable\footnote{Consider an example of a complex symmetric matrix $\S = [1 ~i; ~i ~0]$, which is not eigen-decomposible. See the last paragraph in Section\;\ref{sec:prelim} for a detailed comparison of our work with \cite{ram21}.}. 
How to define graph frequencies when the admittance matrix is not eigen-diagonalizable is left unanswered.


To summarize, in this paper there are three main contributions: 1) A LP is formulated to use existing grid measurement datasets to learn a sparse, complex-valued but Hermitian graph Laplacian matrix offline that represents the electrical inter-dependencies in the grid, ensuring that its eigenvalues are real; 2) A quadratic objective is formulated using a complex GLR to infer the missing grid states online, which can be solved in linear time; and 3) Experimental studies along with comparative studies are conducted on practical 118-bus system to demonstrate the efficacy of the proposed graph learning and grid signal interpolation method. 
\textit{To the best of our knowledge, we are the first to propose the learning of a sparse, complex-valued and Hermitian graph Laplacian matrix with real eigenvalues for the grid state interpolation problem.} 

The paper is organized as follows. Sec. \ref{sec:prelim} consists of fundamental definitions and notations associated with the graph signal processing literature. Sec. \ref{sec:method} details the proposed graph learning problem formulation and transformation into a linear program. Sec. \ref{sec:method2} presents the grid signal interpolation problem formulation and iterative solution computation based on CG. 
Sec. \ref{sec:results} presents the experimental and comparative studies evaluating the proposed algorithm. Finally, the paper is concluded in Sec. \ref{sec:conclusion}.

\section{Graph Spectrum}
\label{sec:prelim}
\subsection{GSP Definitions}

First, we present the notations used in this paper. 
Vectors and matrices are denoted by lower-case (\eg, $\x$) and upper-case (\eg, $\X$) boldface letters, respectively.
$\I_N$ denotes an identity matrix of dimension $N \times N$, $\1_N$ ($\0_N$) denotes an all-one (all-zero) vector of length $N$, and $\0_{N,N}$ denotes an all-zero matrix of dimension $N \times N$. 

Next, we review basic definitions in GSP \cite{ortega18ieee,cheung18}.
Conventionally, an undirected graph $\cG(\cV,\cE,\W)$ has a set $\cV~=~\{1, \ldots, N\}$ of $N$ nodes connected by edges in set $\cE$, where edge $(i,j) \in \cE$ has edge weight $w_{i,j} = W_{i,j}$ specified in symmetric \textit{adjacency matrix} $\W \in \mathbb{R}^{N \times N}$. 
Diagonal \textit{degree matrix} $\D \in \mathbb{R}^{N \times N}$ has diagonal entries $D_{i,i} = \sum_{j} W_{i,j}$. 
Symmetric \textit{combinatorial graph Laplacian matrix} is $\L \triangleq \D - \W$.  
If self-loops $(i,i) \in \cE$ exist, then the symmetric \textit{generalized graph Laplacian matrix} $\cL \triangleq \D - \W + \text{diag}(\W)$ is typically used.
We discuss extension of these definitions to Hermitian graphs with directional edges and complex edge weights next.

\begin{figure}[t]
\centering
\includegraphics[width=0.38\textwidth]{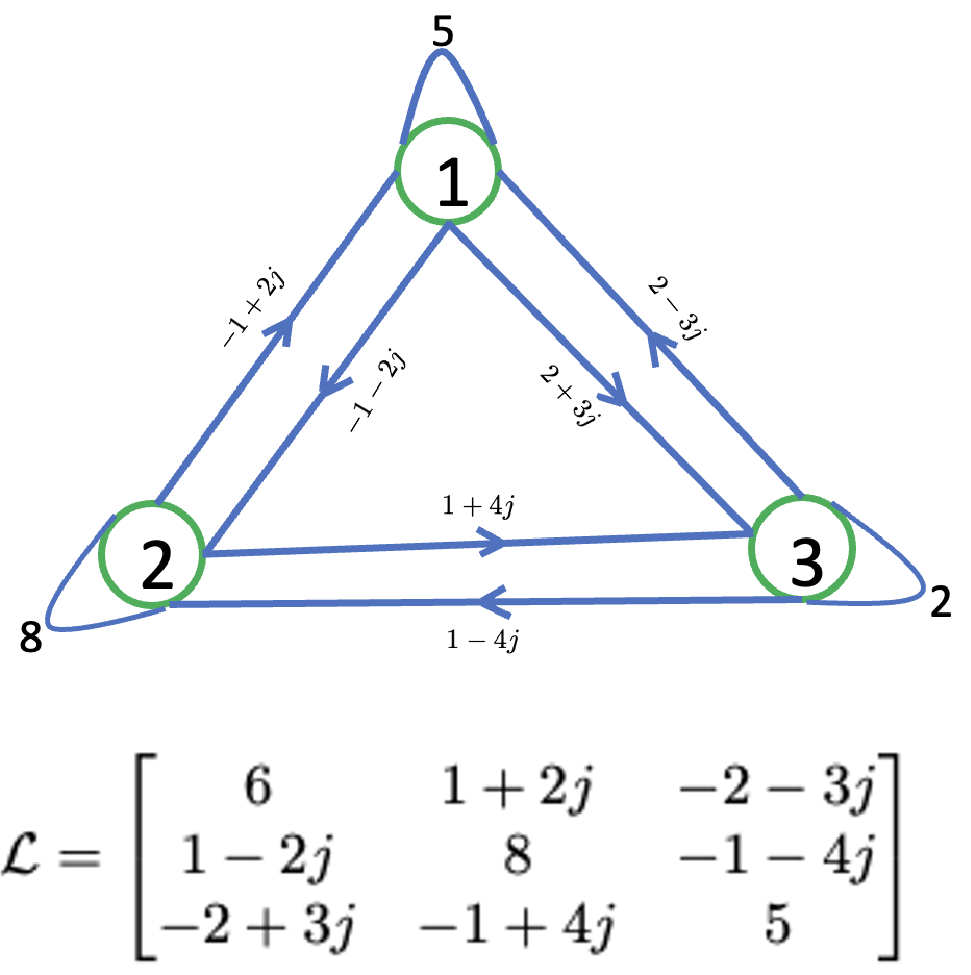}
\vspace{-0.1in}
\caption{An example of a $3$-node Hermitian graph (top)  and its corresponding generalized graph Laplacian $\cL$ (bottom). For each complex weight $w_{i,j} = m_{i,j} e^{j\theta_{i,j}} \in \mathbb{C}$ of a directed edge $[i,j] \in \cE$ (where amplitude is $m_{i,j} \in \mathbb{R}_+$ and phase is $\theta \in [0, 2\pi]$) connecting two nodes $i$ and $j$, the angle $\theta_{i,j}$ describes the conditional mean phase difference between the two phasors, and the amplitude $m_{i,j}$ describes how well the phasors cluster this mean. } 
\label{fig:Her_graph}
\vspace{-10pt}
\end{figure}

\subsection{Hermitian Graph and its Spectrum}

To process complex-valued signal $\x \in \mathbb{C}^N$ (\eg, voltage phasors for $N$ buses), we extend conventional undirected graphs to a directed graph notion called \textit{Hermitian graphs}\footnote{Our Hermitian graph and graph Laplacian definitions differ from existing ones in~\cite{guo2017hermitian, laenen2019directed, abudayah2021hermitian} that focus more narrowly on unweighted graphs. }~\cite{guo2017hermitian, laenen2019directed, abudayah2021hermitian}.
First, each edge $[i,j] \in \cE$ is now \textit{directed} and endowed with a complex-valued edge weight $w_{i,j} \in \mathbb{C}$; the associated edge $[j,i] \in \cE$ in the opposite direction connecting the same two nodes $i$ and $j$ has weight $w_{j,i}$ that is the complex conjugate of $w_{i,j}$, \ie, $w_{j,i} = w_{i,j}^{\text{H}}$.
Each node $i$ may in addition has a real-valued self-loop $w_{i,i} \in \mathbb{R}$. 
This implies that the adjacency matrix $\W \in \mathbb{C}^{N \times N}$ is Hermitian, \ie, $\W = \W^\text{H}$. 
For graphs without self-loops, the corresponding combinatorial graph Laplacian is defined as $\L \triangleq \D - \W$, where $\D \in \mathbb{C}^{N \times N}$ is a diagonal matrix with degree\footnote{This definition of node degree defaults to the undirected graph case when $\W$ is real and symmetric.} $D_{i,i} \triangleq \frac{1}{2} \sum_{j|j\neq i} (w_{i,j} + w_{j,i})$ for node $i$.
Note that $D_{i,i} \in \mathbb{R}$, and thus $\L$ is Hermitian. 
For graphs with self-loops, the corresponding Hermitian generalized graph Laplacian is $\cL \triangleq \D - \W + \text{diag}(\W)$, where node degree is now $D_{i,i} \triangleq w_{i,i} + \frac{1}{2} \sum_{j|j\neq i} (w_{i,j} + w_{j,i})$.

See Fig.\;\ref{fig:Her_graph} for an example of a Hermitian graph $\cG$ with three nodes and the corresponding generalized graph Laplacian $\cL$. 
Note that the diagonal terms of $\cL$ are real, while the off-diagonal terms are in complex conjugate pairs, \ie, $\cL_{i,j} = \cL_{j,i}^\text{H}$.
Weights of directional edges connecting node pairs (complex) and self-loops (real) are also shown. 

A Hermitian graph Laplacian $\L$ (or generalized Laplacian $\cL$) means that the Spectral Theorem \cite{golub12} applies, and eigen-pairs $\{\lambda_k, \v_k\}$ of Hermitian $\L$ (or $\cL$) have real eigenvalues $\lambda_k$'s and mutually orthogonal eigenvectors $\v_k$'s, \ie, $\v_k^\text{H} \v_l = \delta_{k-l}$, where $\delta_{k}$ is the discrete impulse.

Thus, we can define a Hilbert space $\cH$ for complex length-$N$ vectors $\{\x \in \mathbb{C}^N\}$ with an inner product $\langle \u, \v \rangle \triangleq \v^\text{H} \u$ that defines orthogonality---$\{\v_k\}$ is a set of orthogonal basis vectors spanning $\cH$ \cite{vetterli14foundations}. 
A signal $\x \in \mathbb{C}^N$ can be spectrally decomposed to $\balpha = \V^\text{H} \x$, where the $k$-th graph frequency coefficient is  $\alpha_k = \langle \x, \v_k \rangle = \v_k^\text{H} \x$. 
Further, because $\{\v_k\}$ are eigenvectors of Hermitian $\L$, they are successive orthogonal arguments that minimize the \textit{Rayleigh quotient} $\v^\text{H} \L \v$. 
Thus, interpreting $\L$ as the precision matrix of a \textit{Gaussian Markov Random Field} (GMRF) model~\cite{rue2005}, \ie, the probability\footnote{We show in Section\;\ref{sec:com_GLR} that given Hermitian and positive semi-definite (PSD) matrix $\L$, GLR $\x^\text{H} \L \x \in \mathbb{R}_+$ for any $\x \in \mathbb{C}^N$, and thus $\exp(-\x^\text{H} \L \x)$ is within real interval $[0,1]$, and hence can be interpreted as a probability.} $\text{Pr}(\x)$ of signal $\x$ is
\begin{align}
\text{Pr}(\x) \propto \exp \left( - \x^\text{H} \L \x \right).    
\end{align}
This means that $\v_1$ is the most probable signal, $\v_2$ is the next most probable signal orthogonal to $\v_1$, etc. 
Hence, \textit{using GLR $\x^\text{H} \L \x$ as a signal prior means a bias towards low-frequency signals that are also the most probable}.

We contrast our definition of graph frequencies using Hermitian Laplacian $\L$ (or $\cL$) with one in \cite{ram21}, which claimed that a complex symmetric \textit{graph shift operator} (GSO) matrix $\S$ can be eigen-diagonalized as $\S = \U \bLambda \U^\top$ by Theorem 4.4.13 in \cite{horn2012matrix}. 
First, the cited theorem states only that symmetric matrix $\S$ is diagonalizable, \ie, $\S = \Q \bLambda \Q^{-1}$ and $\bLambda$ is a diagonal matrix, iff it is complex orthogonally diagonalizable, \ie, $\S = \U \bLambda \U^\top$. 
It is not clear how graph frequencies are defined if $\S$ is not eigen-diagonalizable.
In such cases, prior $\|\x^\text{H} \S \x\|_2$ would not promote low-frequency signal construction (first eigenvectors of $\S$).
Further, since $\S$ is not Hermitian, prior $\x^\text{H} \S \x$ is not real in general and thus not amenable to fast optimization.

\section{Hermitian Graph Learning}
\label{sec:method}
\subsection{Computing Sparse Complex Precision Matrix as LP}
\label{sec:precision_matrix}

Our first goal is to estimate a sparse complex-valued precision matrix $\P \in \mathbb{C}^{N \times N}$---interpreted as a graph Laplacian matrix $\cL$ to a Hermitian graph $\cG$---from observation matrix $\X = [\x_1, \ldots, \x_K] \in \mathbb{C}^{N \times K}$, $N < K$, where $\x_k \in \mathbb{C}^N$ is the $k$-th zero-mean signal observation (\ie, there is a training dataset composed of $K$ datapoints).
Our formulation generalizes the original CLIME formulation~\cite{cai2011constrained} that seeks a \textit{real} matrix $\P \in \mathbb{R}^{N \times N}$, given \textit{real} empirical covariance matrix $\C \in \mathbb{R}^{N \times N}$.
Specifically, the CLIME formulation is
\begin{align}
\min_{\P} ~~ \left\| \P \right\|_1, ~~
\mbox{s.t.}~ \|\C \P - \I_N \|_{\infty} \leq \rho
\label{eq:clime} 
\end{align}
where $\rho \in \mathbb{R}_+$ is a parameter.
In a nutshell, \eqref{eq:clime} seeks a sparse $\P$ (promoted by the $\ell_1$-norm in the objective) that approximates the \textit{right} inverse\footnote{Note that in general there exist many right inverses $\P ~\mbox{s.t.}~ \C \P = \I$ given positive definite matrix $\C$, while a matrix inverse $\C^{-1}$ is unique.} of $\C$ (enforced by the constraint).
Each $i$-th column $\p_i$ of $\P$ in \eqref{eq:clime} can be solved independently as
\begin{align}
\min_{\p_i} \|\p_i\|_1, ~~
\mbox{s.t.}~ \|\C \p_i - \e_i\|_{\infty} \leq \rho
\label{eq:clime2}
\end{align}
where $\e_i$ is the $i$-th column of $\I_N$---one-hot (canonical) vector with $1$ at the $i$-th entry and $0$ elsewhere. 
\eqref{eq:clime2} can be solved as a \textit{linear program} (LP) using an off-the-shelf LP solver such as \cite{jiang20} with complexity $\mathcal{O}(N^{2.055})$.
The obtained solution $\P$ from \eqref{eq:clime} is not symmetric in general, and \cite{cai2011constrained} post-computes a symmetric approximation $\P^o \leftarrow (\P + \P^\top)/2$.

We generalize CLIME to compute a \textit{complex}-valued $\P \in \mathbb{C}^{N \times N}$ from observation $\X \in \mathbb{C}^{N \times K}$.
We first compute empirical covariance matrix $\C = \frac{1}{N} \X \X^\text{H} \in \mathbb{C}^{N \times N}$. 
Optimization variable $\P$ has real and imaginary parts, $\P = \P^R + j \P^I$, where $\P^R, \P^I \in \mathbb{R}^{N \times N}$.
Our final solution $\P$ must be Hermitian, which implies that its eigenvalues are real by the Spectral Theorem \cite{golub12}. 
This also means that $\P^R$ is symmetric while $\P^I$ is anti-symmetric:
\begin{align}
\P^R = (\P^R)^\top, ~~~
\P^I = - (\P^I)^\top .
\end{align}
Note that anti-symmetry in $\P^I$ means that $P^I_{m,m} = 0, \forall m$. 
Like CLIME, we do not impose (anti-)symmetry on $\P^R$ and $\P^I$ during optimization; instead, given optimized $\P^R$ and $\P^I$, we post-compute final solution as $\P^{R*} \leftarrow (\P^R + (\P^R)^\top)/2$ and $\P^{I*} \leftarrow (\P^I - (\P^I)^\top)/2$ to ensure matrix (anti-)symmetry.

To handle complex $\p_i$, we generalize the objective in \eqref{eq:clime2} $\|\p_i\|_1 = \sum_{n} \|p_{i,n}\|_1$, where $\|p_{i,n}\|_1$ is now the \textit{Manhattan distance}\footnote{Employing the Manhattan distance means that the real $\p_i^R$ and imaginary $\p_i^I$ parts have the same sparsity patterns, which is reasonable given that $\p_i~=~\p_i^R + j\p_i^I$ corresponds to a single node $i$ in the Hermitian graph.} for $p_{i,n} \in \mathbb{C}$, \ie, $\|p_{i,n}\|_1 = |p_{i,n}^R| + |p_{i,n}^I|$.
To retain a linear objective, we define upper-bound auxiliary variables $\bar{\p}^R$ and $\bar{\p}^I$ with linear constraints:
\begin{align}
\bar{\p}^R &\geq \pm \; \p^R_i, ~~~~~
\bar{\p}^I \geq \pm \; \p^I_i
\label{eq:barP_const}
\end{align}
where $\bar{\p}^R \geq \pm \, \p^R_i$ means $\bar{\p}^R \geq \p^R_i$ \textit{and} $\bar{\p}^R \geq - \, \p^R_i$. 
Vector inequality here means entry-by-entry inequality, \eg, $\bar{\p}^R \geq \p^R_i$ means $\bar{p}^R_{n} \geq p^R_{i,n}, \forall n$. 
Thus, the objective becomes
\begin{align}
\sum_{n} \bar{p}^R_{n} + \bar{p}^I_{n} = \1_N^\top \bar{\p}^R + \1_N^\top \bar{\p}^I .
\end{align}

We next generalize the constraint term $\|\C\p_i - \e_i\|_\infty$, where $\|\v\|_\infty = \max_{n} \|v_{n} \|_1$, and $\|v_{n}\|_1$ is the aforementioned Manhattan distance for complex $v_{n}$. 
Expanding $\C\p_i - \e_i$ into its real and imaginary components:
\begin{align}
\C\p_i - \e_i &= (\C^R + j \C^I)(\p^R_i + j \p^I_i) - \e_i 
\label{eq:barS}  \\
&= (\C^R \p^R_i - \C^I \p^I_i - \e_i) + j (\C^R \p^I_i + \C^I \p^R_i) .
\nonumber 
\end{align}
We now define upper-bound auxiliary variables $\bar{\s}^R$ and $\bar{\s}^I$ with corresponding linear constraints:

\vspace{-0.05in}
\begin{small}
\begin{align}
\bar{\s}^R &\geq \pm \left( \C^R \p^R_i - \C^I \p^I_i - \e_i \right), ~~
\nonumber \\
\bar{\s}^I &\geq \pm \left( \C^R \p^I_i + \C^I \p^R_i \right) .
\label{eq:barS_const}
\end{align}
\end{small}\noindent
Thus, the constraint of the optimization $\|\C\p_i - \e_i\|_{\infty} \leq \rho$ becomes
\begin{align}
\bar{\s}^R + \bar{\s}^I \leq \rho \1_N .
\end{align}
Summarizing, the LP formulation to compute complex $\p_i$ is

\vspace{-0.05in}
\begin{small}
\begin{align}
\min_{\p_i, \bar{\p}, \bar{\s}} & \1_N^\top \bar{\p}^R + \1_N^\top \bar{\p}^I, ~ 
\mbox{s.t.} \left\{ \begin{array}{l}
\bar{\p}^R \geq \pm \p^R_i, ~~ \bar{\p}^I \geq \pm \p^I_i \\
\bar{\s}^R \geq \pm (\C^R \p^R_i - \C^I \p^I_i - \e_i) \\
\bar{\s}^I \geq \pm (\C^R \p^I_i + \C^I \p^R_i) \\
\bar{\s}^R + \bar{\s}^I \leq \rho \1_N
\end{array} \right. .
\label{eq:lp}
\end{align}
\end{small}\noindent
Note that if $\C$ is real-valued, then $\C^I = \0_{N,N}$, and $\p^I = \bar{\p}^I = \0_N$ to minimize objective while easing the last constraint to $\bar{\s}^R \leq \rho \1_N$.
Hence, \eqref{eq:lp} defaults to an LP for original CLIME \eqref{eq:clime2} as expected.

\subsection{Complex GLR}
\label{sec:com_GLR}

Having computed $\P^R$ and $\P^I$ via \eqref{eq:lp} (and post-computing $\P^{R*}$ and $\P^{I*}$ to ensure they are symmetric and anti-symmetric, respectively), we interpret Hermitian $\P^* = \P^{R*} + j\P^{I*}$ as the generalized graph Laplacian matrix of a Hermitian graph, \ie, $\cL = \P^*$. 
We show next that the complex GLR $\x^{\text{H}} \cL \x$, where $\x \in \mathbb{C}^N$, is a real-valued quantity, and thus amenable to fast optimization as a quadratic signal prior.

We first note that the theoretical covariance matrix $\tilde{\C}$ (for zero-mean random vector $\x \in \mathbb{C}^N$) is Hermitian:
\begin{align}
\tilde{\C}^{\text{H}} &= \left( \text{E}[\x \x^{\text{H}}] \right)^{\text{H}} = \text{E}[\x \x^\text{H}] = \tilde{\C} .
\end{align}
Thus, by the Spectral Theorem $\tilde{\C}$ has mutually orthogonal eigenvectors $\{\v_k\}$ with real eigenvalues $\{\gamma_k\}$, and we can write positive definite (PD) $\tilde{\C}$ and its inverse $\tilde{\P} = \tilde{\C}^{-1}$ as
\begin{align}
\tilde{\C} &= \sum_k \gamma_k \v_k \v_k^\text{H}, 
~~~~~
\tilde{\P} = \sum_k \lambda_k \v_k \v_k^\text{H}
\label{eq:SLE}
\end{align}
where $\lambda_k = \gamma_k^{-1} \in \mathbb{R}$.
Thus, letting $\cL = \tilde{\P}$,
\begin{align}
\x^\text{H} \cL \x &= \sum_k \lambda_k \x^\text{H} \v_k \v_k^\text{H} \x = \sum_k \lambda_k \|\v_k^\text{H} \x \|^2_2 \in \mathbb{R} .
\label{eq:proofGLR}
\end{align}
Given that our computed precision matrix $\P^*$ is also Hermitian, the same argument in \eqref{eq:proofGLR} can show complex GLR $\x^\text{H} \cL \x$ is also real-valued in practice when $\cL = \P^*$.

\section{Graph-based Interpolation}
\label{sec:method2}
\subsection{Problem Formulation}

Having established graph Laplacian $\cL$, we formulate our complex-valued grid state interpolation problem as follows. 
First, define a \textit{sampling matrix} $\B \in \{0,1\}^{M \times N}$ that selects $M$ entries corresponding to observation $\y \in \mathbb{C}^M$ from sought state vector $\x \in \mathbb{C}^N$, where $M < N$.  
For example, $\B$ that picks out the second and fourth entries from four nodes is
\begin{align}
 \B = \left[ \begin{array}{cccc}
 0 & 1 & 0 & 0 \\
 0 & 0 & 0 & 1
 \end{array} \right]  . 
\end{align}
We formulate a quadratic objective by combining an $\ell_2$-norm fidelity term with our proposed complex GLR as
\begin{align}
\min_{\x} \|\y - \B \x\|^2_2 + \mu \, \x^{\text{H}} \cL \x ,
\label{eq:obj}
\end{align}
where $\mu \in \mathbb{R}_+$ is a real-valued parameter trading off the fidelity term with GLR.
Note that because both $\|\y - \B\x\|^2_2=(\y - \B\x)^{\text{H}} (\y - \B\x)$ and complex GLR $\x^\text{H} \cL \x$ are real, minimization of a real-valued objective \eqref{eq:obj} is well-defined.

The solution $\x^*$ to \eqref{eq:obj} can be obtained by solving the following linear system:
\begin{align}
\left( \B^\top \B + \mu \cL \right) \x^* = \B^\top \y
\label{eq:sol}
\end{align}
where coefficient matrix is $\A = \B^\top \B + \mu \cL$. 
Given $\cL$ computed from \eqref{eq:lp} is PD in general, $\A$ is also PD by Weyl's inequality~\cite{merikoski2004}.
Because $\A$ is sparse, Hermitian and PD, $\x^*$ in \eqref{eq:sol} can be obtained without matrix inversion using \textit{conjugate gradient} (CG) \cite{hestenes1952methods} in roughly linear time.

\subsection{Summary of Proposal}
The proposed graph learning and interpolation steps are summarized in Alg.\;\ref{alg:alg1}. 
First, $K$ signal observations $\{\x_k\}$ are used to compute empirical covariance matrix $\C = \frac{1}{N} \X \X^\text{H}$. 
Then, the LP in \eqref{eq:lp} is solved $N$ times to obtain $N$ columns $\{\p_i\}$ of precision matrix $\P$ using an off-the-shelf solver such as \cite{jiang20}. 
We set graph Laplacian $\cL$ to be $\P^* = \P^{R*} + j \P^{I*}$, where $\P^{R*}$ and $\P^{I*}$ are post-processed to be symmetric and anti-symmetric, respectively.  
The sampling matrix $\B$ is set based on the samples that are available to be used for interpolation. 
Finally, CG \cite{hestenes1952methods} is used to solve the linear system in \eqref{eq:sol} to obtain all grid states $\x^*$. 

\begin{algorithm}[htp!]
 \caption{Hermitian graph-based state interpolation}
 \label{alg:alg1}
 \textbf{Input:} observation dataset $\X \in \mathbb{C}^{N \times K}$, parameters $\rho, \mu \in \mathbb{R}_+$
\\
\textbf{Initialization:}
\\
$\C\gets \frac{1}{N} \X \X^\text{H}$, $i\gets 1$.
\\
\textbf{Algorithm:}
\\
\textbf{while} $i\leq N$ \ \textbf{do}
\begin{itemize}
\item Solve LP in \eqref{eq:lp} to compute $\p_i$.
\end{itemize}
\textbf{end while} \\
$\P\gets [\p_1 \ldots \p_N]$. 
\\
$\P^{R*} \gets (\P^R+(\P^R)^\top)/2$, ~~$\P^{I*} \gets (\P^I-(\P^I)^\top)/2$. \\
$\cL \gets \P^* = \P^{R*} + j \P^{I*}$.
\\
Solve linear system in Eq. \eqref{eq:sol} via CG \cite{hestenes1952methods} to compute $\x^*$.
\end{algorithm}

\section{Experiments}
\label{sec:results}
\begin{table*}[t]
\centering
\caption{Average MSE with 95\% Confidence Interval for the reconstruction of voltage magnitude using different set of samples}
\begin{tabular}{|c|c|c|c|}
\hline
\begin{tabular}[c]{@{}c@{}}Number of samples\end{tabular} & Scheme 1 & Scheme 2 & Proposed \\ \hline
30                                                                 & 0.0231 $\pm$ 0.0063               & 0.0241 $\pm$ 0.0109              & 0.0229 $\pm$ 0.068    \\ \hline
40                                                                    & 0.0215 $\pm$ 0.0047                & 0.0231 $\pm$ 0.0063              & 0.0216 $\pm$ 0.0051    \\ \hline
50                                                                    & 0.0165$ \pm$ 0.0035               & 0.0194 $\pm$ 0.0062              & 0.0168$ \pm$ 0.0037    \\ \hline
60                                                                   & 0.0154 $\pm$ 0.0048               & 0.0162 $\pm$ 0.0026              & 0.0153 $\pm$ 0.0047    \\ \hline
70                                                                    & 0.0140 $\pm$ 0.0026               & 0.0149 $\pm$ 0.0022              & 0.0143 $\pm$ 0.0028    \\ \hline
80                                                                    & 0.0129 $\pm$ 0.0022                & 0.0144 $\pm$ 0.0017              & 0.0129 $\pm$ 0.0019    \\ \hline
90                                                                    & 0.0118$ \pm$ 0.0018                & 0.0128 $\pm$ 0.0030              & 0.0116$ \pm$ 0.0017    \\ \hline
100                                                                   & 0.0103 $\pm$ 0.0011               & 0.0110 $\pm$ 0.0015              & 0.0102 $\pm$ 0.0010    \\ \hline
\end{tabular}
\label{tab:MSE_mag}
\vspace{-10pt}
\end{table*}

\begin{table*}[t]
\centering
\caption{Average MSE with 95\% Confidence Interval for the reconstruction of voltage phase angles using different set of samples}
\begin{tabular}{|c|c|c|c|}
\hline
\begin{tabular}[c]{@{}c@{}}Number of samples\end{tabular} & Scheme 1 & Scheme 2 & Proposed \\ \hline
30                                                                    & 0.0860 $\pm$ 0.0248                & 0.0911 $\pm$ 0.0257              & 0.0870 $\pm$ 0.0255    \\ \hline
40                                                                    & 0.0814 $\pm$ 0.0235                & 0.0891 $\pm$ 0.0248              & 0.0811 $\pm$ 0.0234    \\ \hline
50                                                                    & 0.0765$ \pm$ 0.0221               & 0.0864 $\pm$ 0.0244              & 0.0771$ \pm$ 0.0226    \\ \hline
60                                                                   & 0.0722 $\pm$ 0.0227              & 0.0790 $\pm$ 0.0239              & 0.0720 $\pm$ 0.0224    \\ \hline
70                                                                    & 0.0670 $\pm$ 0.0176                & 0.0775 $\pm$ 0.0220              & 0.0669 $\pm$ 0.0180    \\ \hline
80                                                                    & 0.0640 $\pm$ 0.0163               & 0.0704 $\pm$ 0.0225              & 0.0642 $\pm$ 0.0169    \\ \hline
90                                                                    & 0.0580$ \pm$ 0.0164                & 0.0670 $\pm$ 0.0180              & 0.0585$ \pm$ 0.0166    \\ \hline
100                                                                   & 0.0531 $\pm$ 0.0150                & 0.0590 $\pm$ 0.0157              & 0.0528 $\pm$ 0.0152    \\ \hline
\end{tabular}
\label{tab:MSE_phase}
\end{table*}

\begin{figure*}[t]
\centering
\includegraphics[width=0.4\textwidth]{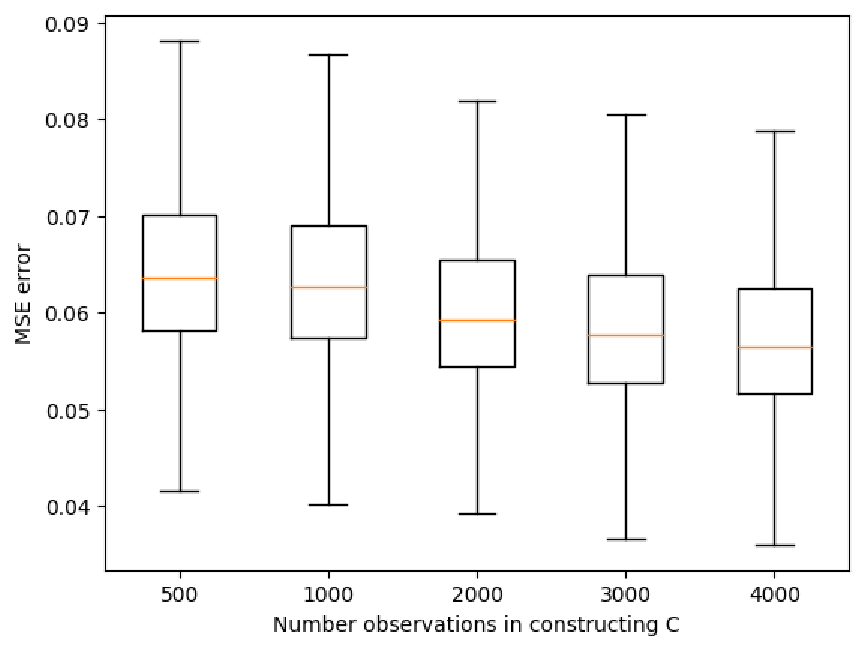}
\vspace{-0.1in}
\caption{Reconstruction of voltage phase angles under matrix $\C$ constructed from different number of observations} 
\label{fig:box_plot}
\vspace{-10pt}
\end{figure*}

\begin{table}[]
\centering
\caption{Average Execution time (S) for Interpolation}
\begin{small}
\begin{tabular}{|c|c|}
\hline
Methods            & Total Execution Time \\ \hline
Scheme 1 & 1.925                        \\ \hline
Scheme 2 & 2.082                       \\ \hline
Proposed           & 1.596                       \\ \hline
\end{tabular}
\end{small}
\label{tab:time}
\vspace{-15pt}
\end{table} 

In this section, we conduct practical experiments to evaluate the performance of our proposed grid state interpolation method on the IEEE 118 bus system. 15,000 observations are generated using MATPOWER that is run in MATLAB. The power demands are sampled uniformly in the range of [80\% to 120\%] of the default demand values as is common practice in existing literature \cite{Fioretto2020,pan2020}. The generation mix considered includes renewables and traditional bulk generation sources. Wind and solar power sources compose of 8.4\% and 2.3\% of overall generation in the United States \cite{eia2020} and the same proportion is applied in our simulations. Each wind farm is assumed to be composed of turbines representing a total of 1800 m$^2$ swept area. Each solar farm consists of 10,000 photovoltaic panels (1.67 m$^2$ per panel). 
Uncertainties in wind speed and solar generation are modeled using the Weibull probability distribution with the shape and location parameters taking values of (2, 5) and (2.5, 6), respectively, in a manner similar to reference \cite{Wang2022}. Each observation is composed of the complex voltage phasor of 118 buses. Alg. \ref{alg:alg1} is implemented on a MacBook Pro Apple M1 Chip with 8-core CPU, 8-core GPU, 16-core Neural Engine, and 16GB unified memory.

In order to compute the Hermitian matrix $\P$ as outlined in Sec. \ref{sec:method}, 5000 observations are randomly selected to compute the empirical covariance matrix $\C$. The remaining 10000 observations are utilized to test the interpolation of missing grid states. Then, in the interpolation of the missing grid states, we randomly select $30$, $40$, $50$, $60$, $70$, $80$, $90$, $100$ samples out of $118$ samples to reconstruct the remaining missing grid states (\ie, if $30$ samples are utilized, then $88$ samples are inferred/interpolated by solving the unconstrained quadratic program listed in Sec. \ref{sec:method2}. The mean square error of the inferred grid states from the ground truth is computed for the 10,000 samples along with the 95\% confidence interval and listed in Tables \ref{tab:MSE_mag} and \ref{tab:MSE_phase} for voltage magnitude and phase angles respectively. It is clear that our proposal has very low error margins (i.e. within 2.3\% even when only 25\% of grid measurements are available for interpolating the remaining state variables. This is significantly more efficient than traditional state estimation processes that require as many samples as the number of states being inferred. 

We examine the effect of varying the number observations in constructing the $\C$ matrix on the accuracy of the interpolation of missing grid states. Intuitively, one can expect the more observations that included, the more reflective will the learnt $\P$ matrix will be of the electrical interdependencies in the system. The number of samples is fixed to be 100, and the MSE error of state interpolation is tested with different number of observations in constructing $\C$. 
The result is shown in Fig.\;\ref{fig:box_plot}, in which the average MSE error gets lower as the number of observations rise. 
As matrix $\P$ is computed once offline, the computational overhead incurred will not affect the interpolation processes. In Table \ref{tab:time}, we list the average total execution time for the interpolation algorithm. We are able to compute the missing grid states within 1.5 seconds. This allows for near real-time inferences of grid states for a system as large as the IEEE 118 bus system. Our approach is conducive for near real-time monitoring of the entire grid from a small set of sample states.

Finally, we conduct comparative studies with two GSP based schemes. The first scheme is an interpolation method regularized using  \textit{graph total variation} (GTV) \cite{ram21}, resulting in an $\ell_2$-$\ell_1$-norm minimization objective. 
We minimize it using a known primal-dual algorithm~\cite{chambolle2011first}. 
For the second scheme, instead of learning complex-valued $\cL$, we learned two Laplacian matrices for real and imaginary components of the signals separately, and then performed interpolation via \eqref{eq:sol}. The proposed method reduced the MSE by 7.68\%  and 9.86\% on average for voltage magnitude and phase angle respectively in comparison to Scheme 2, while the proposed method is comparable to Scheme 1. Average execution times are shown in Table~\ref{tab:time}. 
The proposed method is noticeably faster than both competing schemes: by 17\% and 18\%, respectively. 
Furthermore, as the $\P$ matrix is Hermitian, we can ensure stable computation, while reference \cite{ram21} utilizes the nodal admittance matrix that may have complex eigenvalues and may not be eigen-decomposible. Our proposal also does not require knowledge of the underlying system topology and parameters.


\section{Conclusion}
\label{sec:conclusions}
\label{sec:conclusion}
In this paper, we have presented a novel GSP based approach to interpolate a small number of grid measurements samples to compute the voltage magnitude and phase angles of all buses in the grid from which other grid states such as real and reactive power flows can be computed. Our proposal does not require knowledge of the underlying grid structure and electrical parameters. Further, we have demonstrated in our experiments that significantly smaller number of samples (e.g. only 25\% of grid states can be utilized to reconstruct all grid states) are required to effectively interpolate grid states in comparison to traditional state estimation techniques.  
We demonstrate the performance of our proposal in terms of accuracy and computational time by comparing our scheme with another GSP approach proposed recently in the literature and a primal-dual algorithm based method for solving the formulated optimization problems. 



\begin{footnotesize}
\bibliographystyle{IEEEbib}
\bibliography{ref2}
\end{footnotesize}

\end{document}